# The Sun-Like Activity of the Solar Twin 18 Scorpii

Jeffrey C. Hall,[1] Gregory W. Henry,[2] & G. Wesley Lockwood[1]

[1]Lowell Observatory, 1400 West Mars Hill Road, Flagstaff, AZ 86001

[2]Tennessee State University, Center of Excellence in Information Systems, 3500 John A. Merritt Blvd., Box 9501, Nashville, TN 37209

## Abstract

We present the results of ten years of complementary spectroscopic and photometric observations of the solar twin 18 Scorpii. We show that over the course of its ~7 year chromospheric activity cycle, 18 Sco's brightness varies in the same manner as the Sun's, and with a likely total brightness variation of 0.09%, similar to the 0.1% decadal variation in the total solar irradiance.

1. Introduction

Since the advent of space-borne solar observatories, the Sun's total brightness has been found to vary directly with its activity level at a level of about 0.1% (Fröhlich & Lean 1998). It also appears to evolve on century (Lockwood & Stamper 1999) and longer (Rind 2002) timescales, though debate persists regarding the presence or absence of a secular increase between the 1986 and 1996 minima (Fröhlich & Lean 1998, Willson 1997). The decadal forcing of this observed total solar irradiance (TSI) variation appears unlikely to have been a dominant driver of global warming since 1970 (Wang, Lean, & Sheeley 2005), but longer-term (and in some cases short-term) signals undeniably exist in the climate record (Rind 2002). The pronounced decadal variations of solar activity in the UV may have important effects on tropospheric conditions through modulation of ozone (Shindell et al. 1999, Haigh 2004).

To improve our perspective on our own star and its variability, we may turn to the most nearly Sun-like stars as additional examples of ostensibly typical stellar behavior. The Fraunhofer H and K lines of singly ionized calcium are well known as sensitive proxies for the chromospheric manifestations of the Sun's activity cycle (White & Livingston 1981, White et al. 1998) as well as the cycles of Sun-like stars (Wilson 1978, Baliunas et al. 1995). In addition to their chromospheric activity, true solar twins are required to be "nearly indistinguishable" from the Sun in their gross physical parameters (Cayrel de Strobel 1997), and they have been elusive. Only one star accessible to extant long-term programs, the fifth-magnitude star 18 Scorpii, has met all the criteria (de Mello

& da Silva 1997), though additional good candidates have been proposed (Soubiran & Triaud 2004).

To study the cyclic and irradiance behavior of Sun-like stars, long-term (i.e., decadal or longer), synoptic observations are necessary. Such studies have revealed important clues about the irradiance variability of Sun-like stars of different ages (Radick et al. 1998), but since concerted searches for solar analogs were simultaneous with (or post-dated) the long-term studies, most of the resulting best candidates have not been observed in detail. To remedy this, we have been observing a large set of the closest solar analogs and twins both spectroscopically in Ca II H&K (Hall & Lockwood 1995, Hall, Lockwood, & Skiff 2007) and photometrically in the *b* and *y* bandpasses of the Strömgren system (Henry 1999) for over ten years. In this paper, we demonstrate that the solar twin 18 Scorpii exhibits luminosity variations remarkably similar to those of the Sun over the course of its activity cycle, lending confidence – to the extent that this single data point permits – in the use of Sun-like stars as reliable proxies for the likely envelope of solar behavior on decadal to millennial timescales.

2. Observations and Data Analysis

The spectroscopic observations presented herein were obtained with the Solar-Stellar Spectrograph (SSS), operated at a 1.1-meter telescope at Lowell Observatory's dark sky site near Flagstaff, AZ. The spectrograph has been in regular operation for 12 years, and allows observations both of Sun-like stars, as well as the unresolved Sun itself, via dual fiber optic feeds to the spectrograph input. We collect spectra of the Ca II H&K region

($\lambda\lambda$3860-4010 Å) on a charge coupled device (CCD) detector and reduce the data with routines written in the Interactive Data Language (IDL). We measure the Ca II H&K emission in 1 Å rectangular band passes centered on the line cores, and we express our results in terms of the excess flux $\Delta\mathcal{F}_{HK}$, the fraction of the emission measured in the 1 Å band pass arising from dynamo-related magnetic activity (see, e.g., Schrijver, Dobson, & Radick 1989). Details of the SSS facility, the data reduction procedure, as well as how the measured emission in the H and K line cores is converted to $\Delta\mathcal{F}_{HK}$, have been presented by Hall, Lockwood, & Skiff 2007.

We acquired the brightness measurements of 18 Scorpii with the T8 0.8 m automatic photometric telescope (APT) at Fairborn Observatory, located in the Patagonia Mountains of southern Arizona (Henry 1999). The T8 APT is equipped with a two-channel precision photometer employing two EMI 9124QB bi-alkali photomultiplier tubes to make simultaneous measurements in the Strömgren *b* and *y* pass bands. The APT measures the difference in brightness between a program star and a nearby constant comparison star or stars with a typical precision of 0.13% rms for bright stars ($V < 8.0$). We observed three comparison stars along with 18 Scorpii: HD 143841 ($V = 7.12$, $B-V = 0.43$), HD 144892 ($V = 6.69$, $B-V = 0.51$), and HD 141128 ($V = 7.00$, $B-V = 0.47$). Intercomparison of the various differential brightnesses between the three comparison stars showed that HD 143841 and HD 144892 exhibited the greatest constancy; the standard deviation of their night-to-night differential brightnesses averaged only 0.12% over the 10 year timespan of the observations. Seasonal means of the HD 143841 vs. HD 144892 differential brightnesses scattered about their grand mean by only 0.02%. To maximize the precision of our 18 Scorpii differential brightnesses, we chose to combine the Strömgren *b* and *y* differential brightnesses into a single (*b*+*y*)/2 pass band and also to compute 18 Scorpii's differential brightnesses with respect to the *mean* of the two best comparison stars. As shown below, this allowed us resolve subtle brightness variations in 18 Scorpii above the combination of measurement errors and any low-level intrinsic variability in the two comparison stars. Additional information on the operation of the telescope and photometer, observing procedures, and data reduction techniques can be found in Henry (1999).

3. Results

In Figure 1 and Table 1 we present the spectroscopic and photometric behavior of 18 Scorpii from 1997 through 2006. All data points are seasonal means of typically ~60 APT observations and ~20 SSS observations per season; error bars show the standard deviation of the mean. For ease of comparison with the familiar 0.1% TSI variation over the solar activity cycle, we have converted all flux measurements and photometric magnitudes to percent variations about their respective means.

The SSS observations of 18 Sco are shown as blue diamonds in panel A of Figure 1, wherein we show the variation of each seasonal mean about the grand mean of the means. For comparison, the SSS seasonal mean $\Delta\mathcal{F}_{HK}$ for the Sun from 1994 through 2006 are also shown, as open purple squares. The Fairborn differential photometric data appear in panel B, shown as the difference in brightness between each seasonal mean and the grand. The orange and yellow data points are the individual series for 18 Sco relative to the comparison stars HD 143841 and HD 144892, respectively. The red data series is the weighted mean of these two, and is offset by 0.2% for clarity.

The data in Figure 1A show that 18 Sco exhibits a clear chromospheric activity cycle that peaked in 2001 and reached a pronounced minimum in 2004. In the 2005 and 2006 observing seasons, the SSS observations showed a surprisingly sharp rise in 18 Sco's activity, exceeding the full amplitude of the previous cycle in just two years. As is demonstrated by the overlaid solar data, the amplitude of the most recent cycle in 18 Sco relative to its mean is comparable (so far) to that of the Sun. Its absolute level of chromospheric activity, however, is somewhat greater than the Sun's; in terms of the dimensionless index *S* used by the Mount Wilson program to characterize stellar activity (Baliunas et al. 1995), our mean derived value for the Sun is 0.171, while for 18 Sco it is 0.182.

The photometric data for 18 Sco also vary directly with its activity cycle, with amplitude in (*b+y*)/2 of 0.12%. A Spearman rank-order test applied to the 10-year data set yields $\rho = 0.620$, which for 10 data pairs rejects the null hypothesis of non-correlated data sets at the ~95% confidence level. Slightly elevated variability is apparent in the records for the comparison stars from 1999-2001, and this may be contaminating the differential photometry in that time period (evident in the more variable character and larger error bars for those seasons in Figure 1B); possibly as a result of this there is no significant correlation between the spectroscopy and photometry for the seasons between 1997 and 2000. However, since 2001, the two comparison stars appear to have been extremely stable, and for the six seasonal means between 2001 and 2006, we obtain $\rho = 0.943$, rejecting the null hypothesis at a confidence level of 99.5%.

We do not expect, however, that variability in the Strömgren *b* and *y* pass bands should be equal to the variations in a star's *total* brightness. Two independent assessments find that *b* and *y* variability should be larger than TSI variability by factors of 1.34 (Radick et al. 1998) and 1.39 (Fligge et al. 1998). Reducing the 0.12% amplitude in $\Delta(b+y)/2$ for 18 Sco by the mean of these estimates yields a likely total brightness variation of ~0.09%, very similar to that observed for the Sun.

3. Conclusions

Reconstructions of the evolution of solar luminosity, especially in regard to its behavior during grand minima, can be usefully guided by these stellar observations, and this has led to numerous efforts to characterize the nature of Sun-like stars, both in terms of their general parameters (e.g., Henry et al. 1996, Gray et al. 2003) and their long-term variability (e.g., Baliunas et al. 1995, Radick et al. 1998). Chromospheric proxies such as Ca II H&K have been fruitful since they are easily observed and display a marked variation for cycles comparable to the Sun's. It is essential, however, to also observe the brightness cycles, which are more closely tied to a star's climate forcing ability on such planets as it may have than are the evanescent HK variations. We know that the direct variation of TSI with Ca H&K emission observed for the Sun holds in general for solar age stars (Radick et al. 1998), but the targets in that study are generally not "nearly indistinguishable" from the Sun. The present results demonstrate for the one star that does seem to fit the criterion, both the chromospheric activity cycle and the brightness variations very closely mimic those of the Sun for the 10-year span of observations accumulated thus far.

As we begin to observe a few stars that appear to have made significant transitions between cycling and non-cycling states (Hall, Lockwood, & Skiff 2007), analysis of their complementary photometric behavior should further constrain our understanding of the likely envelope of luminosity excursions of Sun-like stars; these results are in preparation. The next two to three observing seasons will also likely reveal the full amplitude of the newly emerging and rather abruptly rising cycle in 18 Sco, which already exceeds that of the 1996-2004 cycle.

J. C. H. and G. W. L. acknowledge support from grant ATM-04407159 from the National Science Foundation. G. W. H. acknowledges support from NASA grant NCC5-511 and NSF grant HRD-9706268.

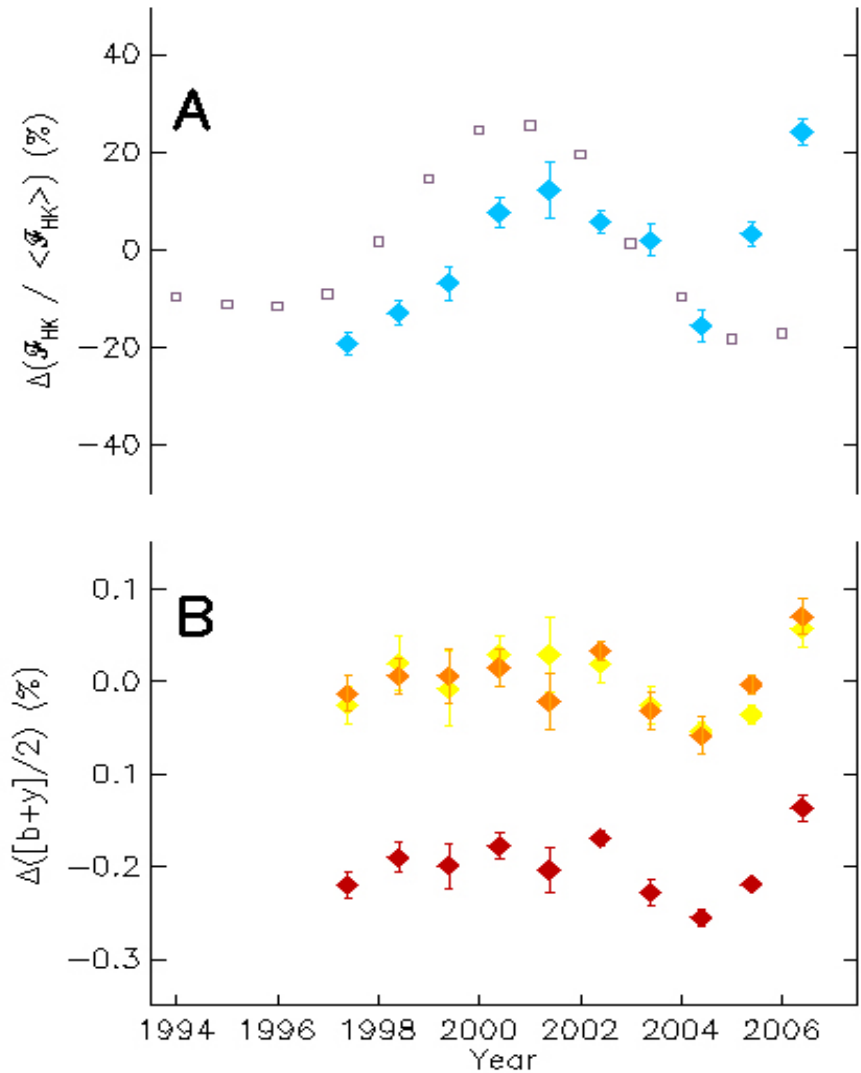

Figure 1. The photometric and spectroscopic behavior of 18 Scorpii, 1997-2006. In panel A, the variation of the Ca II H&K magnetic flux about its mean is shown by blue diamonds. Our thirteen years of observations of the solar HK variation (purple squares) are shown for comparison, scaled the same way. In panel B, we present the photometric light curves. Brightness increases upward. Orange and yellow diamonds show the individual variations relative to two different comparison stars. The red series is the weighted mean of the first two, and is offset by 0.2% for clarity.

Table 1.  Spectroscopic and photometric variations in 18 Scorpii from 1997 through 2006.  For each year we show the seasonal mean and the standard deviation of the mean for the Ca II HK flux relative to its 10-year mean (column 2) and the brightness of 18 Sco relative to its 10-year photometric mean (column 3).  The Ca II H&K variations in column 2 correspond to the blue diamonds in Figure 1, and the ($b+y$/2) variations in column 3 correspond to the red diamonds, which are the weighted mean of the two individual series shown in yellow and orange.

| Year | $\Delta\mathcal{F}_{HK}$ / $\langle\Delta\mathcal{F}_{HK}\rangle$ (%) | $\Delta([b+y]/2)$ (%) |
|---|---:|---:|
| 1997 | -19 ± 2 | -0.018 ± 0.001 |
| 1998 | -13 ± 2 | 0.009 ± 0.002 |
| 1999 | -7 ± 3 | 0.000 ± 0.003 |
| 2000 | 7 ± 3 | 0.018 ± 0.002 |
| 2001 | 12 ± 6 | 0.000 ± 0.003 |
| 2002 | 5 ± 2 | 0.028 ± 0.001 |
| 2003 | 1 ± 3 | -0.028 ± 0.002 |
| 2004 | -16 ± 3 | -0.055 ± 0.001 |
| 2005 | 3 ± 2 | -0.018 ± 0.001 |
| 2006 | 24 ± 2 | 0.064 ± 0.002 |